\documentclass[twocolumn,           
               showpacs,            
               preprintnumbers,     
               aps,                 
               prl,          	    
               letterpaper,             
               superscriptaddress,      
               nofootinbib,         
               tightenlines,        
               floats,floatfix      
               ]{revtex4-1}
\usepackage{graphicx}  
\usepackage{dcolumn}   
\usepackage{bm}        
\usepackage{amsmath,amssymb}
\usepackage{color}
\usepackage{hyperref}

\begin{document}
\title{Cosmological signatures of ultralight dark matter with an axionlike potential}
\author{Francisco X. Linares Cede\~no}
\email{fran2012@fisica.ugto.mx}
\affiliation{
Departamento de F\'isica, DCI, Campus Le\'on, Universidad de
Guanajuato, 37150, Le\'on, Guanajuato, M\'exico}
\author{Alma X. Gonz\'alez-Morales}
\email{alma.gonzalez@fisica.ugto.mx}
\affiliation{
Departamento de F\'isica, DCI, Campus Le\'on, Universidad de
Guanajuato, 37150, Le\'on, Guanajuato, M\'exico}
\affiliation{
Consejo Nacional de Ciencia y Tecnolog\'ia, Av. Insurgentes Sur 1582. Colonia Cr\'edito Constructor, Del. Benito Ju\'arez, C.P. 03940, M\'exico D.F. M\'exico}
\author{L. Arturo Ure\~na-L\'opez}
\email{lurena@ugto.mx}
\affiliation{
Departamento de F\'isica, DCI, Campus Le\'on, Universidad de
Guanajuato, 37150, Le\'on, Guanajuato, M\'exico}
   
\date{\today}
\begin{abstract}
Nonlinearities in a realistic axion field potential may play an important role in the cosmological dynamics. In this paper we use the Boltzmann code CLASS to solve the background and linear perturbations  evolution of an axion field and contrast our results with those of CDM and the free axion case. We conclude that there is a slight delay in the onset of the axion field oscillations when nonlinearities in the axion potential are taken into account. Besides, we identify a tachyonic instability of linear modes resulting in the presence of a bump in the power spectrum at small scales. Some comments are in turn about the true source of the tachyonic instability, how the parameters of the axionlike potential can be constrained by Ly-$\alpha$ observations, and the consequences in the stability of self-gravitating objects made of axions.\end{abstract}
 
\maketitle

\section{Introduction}
Modern cosmological observations have brought about a large amount of data \cite{Ade:2015xua,Albareti:2016xlm}, making it possible to constrain, with high accuracy, theoretical models describing the Universe at large scales. The so-called Lambda Cold Dark Matter ($\Lambda$CDM) paradigm is very successful at reproducing cosmological observations \cite{Ade:2015xua} but it requires a dark matter (DM) component ($\approx 26\%$), effectively described by collisionless particles that interacts mostly gravitationally with other matter components\cite{Liddle:1993fq,Eisenstein:1997jh,Peebles:2017bzw}. However, there are longstanding discussions about how well the $\Lambda$CDM model describes the Universe at galactic and sub-galactic scales\cite{Weinberg:2013aya,Garrison-Kimmel:2014kia}. The solution to these problems may come from the specific properties of the DM, or from an interplay between the DM properties and kinematic effects with baryons, but still the incompleteness of galactic observations may impair our ability to infer the DM distribution properties from them. Given the current status, the further development of theoretical models still is very much desirable if one is to elucidate the properties of this matter component of the Universe.

According to recent studies, axion DM has become a compelling candidate to replace CDM\cite{Hui:2016ltb,Marsh:2015xka,Magana:2012ph}, and even some experiments have been already set up to have a direct detection of this elusive particle \cite{Asztalos:2009yp,Avignone:1997th,Bernabei:2001ny,Morales:2001we,Zioutas:2004hi,Bradley:2003kg}. In particular, there are several proposals for detection of ultralight axions (ULA) using laser interferometers \cite{Aoki:2016kwl}, analyzing the frequency and dynamics of pulsars \cite{Khmelnitsky:2013lxt,Blas:2016ddr}, and also in gravitational wave detectors \cite{Aoki:2017ehb}. Nonetheless, there are still many open questions such as what is the right axion mass limits one can place by using, for instance, galactic kinematics\cite{Urena-Lopez:2017tob,Gonzales-Morales:2016mkl,Chen:2016unw,Calabrese:2016hmp,Marsh:2015wka}, and Lyman-$\alpha$ observations\cite{Irsic:2017yje,Armengaud:2017nkf}. At the cosmological level, axion models have been studied considering it as a free scalar field, i.e., as a scalar field endowed with a quadratic potential $V(\phi)=m^2 \phi^2/2$ \cite{Urena-Lopez:2015gur,Marsh:2010wq,Marsh:2012nm,Marsh:2013ywa,Hlozek:2014lca,Marsh:2011bf}. However, a more realistic form of the axion potential is the trigonometric one,
\begin{equation}
  V(\phi) = m^2_\phi f^2 \left[ 1+\cos\left( \phi/f \right) \right] \, ,   \label{eq:0}
\end{equation}
where $m_\phi$ is the axion mass and $f$ is the decay constant of the axion. The axion potential~\eqref{eq:0} originally arose in QCD with the aim to solve the strong CP problem \cite{Peccei:1977ur,Peccei:1977hh,Weinberg:1977ma,Wilczek:1977pj}, and the potential of such field arises from non-perturbative effects which generate a periodic behavior after the breaking of the Peccei-Quinn symmetry $U(1)_{PQ}$ due to instantons\cite{Sikivie:2006ni,Cheng:2001ys,Dine:1982ah}. More recently, it has been argued that axions emerge in string theories from the compactification of extra dimensions\cite{Svrcek:2006yi,Olsson:2007he,Arvanitaki:2009fg}. 

Given the motivations above, our aim in this paper is to study the axion field as source of DM with its corresponding trigonometric potential~\eqref{eq:0}. For that purpose, we present, for the first time, an analysis of the cosmological evolution, from radiation domination up to the present day, of an axion field taking into account the whole properties of the potential~\eqref{eq:0}. This is accomplished by: $\left.1\right)$ transforming the standard cosmological equations for both, the background and the linear perturbations into a dynamical system, and $\left.2\right)$ using an amended version of the Boltzmann code CLASS\cite{Lesgourgues:2011re} to obtain accurate numerical solutions.
We analyze the differences in the linear process of structure formation of the axion field with respect to the free (quadratic potential) and the CDM cases. For the sake of concreteness we present all the results for a fiducial model with axion mass $m_\phi=10^{-22}$ eV, but we have verified that the qualitative features hold for other masses in the range, $10^{-26}< m_\phi/$eV$< 10^{-20}$.

\section{Background Dynamics \label{sec:background-dy}}
The equations of motion for a scalar field $\phi$ endowed with the potential~\eqref{eq:0}, in a homogeneous and isotropic space-time with null spatial curvature, are given by
\begin{subequations}
\label{eq:2}
  \begin{eqnarray}
    H^2 &=& \frac{\kappa^2}{3} \left( \sum_j \rho_j +
      \rho_\phi \right) \, ,\quad \dot{\rho}_j = - 3 H (\rho_j + p_j ) \,
    , \\
    \dot{H} &=& - \frac{\kappa^2}{2} \left[ \sum_j (\rho_j +
      p_j ) + (\rho_\phi + p_\phi) \right] \, , \label{eq:2a} \\
    \ddot{\phi} &=& -3 H \dot{\phi} + m_\phi^2 f \sin(\phi/f)  \, , \label{kge}
  \end{eqnarray}
\end{subequations}
where $\kappa^2 = 8\pi G$, $\rho_j$ and $p_j$ are the energy and pressure density of ordinary matter, a dot denotes derivative with respect to cosmic time $t$, and $H = \dot{a}/a$ is the Hubble parameter. The scalar field energy density and pressure are given by the canonical expressions $\rho_\phi = (1/2)\dot{\phi}^2+V(\phi)$ and $p_\phi = (1/2)\dot{\phi}^2-V(\phi)$.

We define a new set of polar coordinates as in \cite{Copeland:1997et,Urena-Lopez:2015gur,Urena-Lopez:2015odd}, 
\begin{subequations}
\label{eq:backvars}
\begin{gather}
\frac{\kappa \dot{\phi}}{\sqrt{6} H}  \equiv  \Omega^{1/2}_\phi \sin(\theta/2), \quad \;
 \frac{\kappa V^{1/2}}{\sqrt{3} H} \equiv \Omega^{1/2}_\phi \cos(\theta/2) \, , \\
y_1  \equiv -2\ \sqrt{2} \, \frac{\partial_{\phi}V^{1/2}}{H} \, ,
\end{gather}
\end{subequations}
with which the Klein-Gordon equation~\eqref{kge} takes the form of the following dynamical system:
\begin{subequations}
\label{eq:new4}
  \begin{eqnarray}
  \theta^\prime &=& -3 \sin \theta + y_1 \, ,\quad  \Omega^\prime_\phi = 3 (w_{tot} + \cos\theta)
  \Omega_\phi \label{eq:new4c} \, ,\\
  y^\prime_1 &=& \frac{3}{2}\left( 1 + w_{tot} \right) y_1 + \frac{\lambda}{2}\Omega_{\phi}\sin \theta \,
  , \label{eq:new4b}
\end{eqnarray}
\end{subequations}
where $\lambda = 3/\kappa^2 f^2$ and $\Omega_\phi = \kappa^2 \rho_\phi/3H^2 $ is the standard scalar field density parameter. Here, a prime denotes derivative with respect to the number of $e$-foldings $N \equiv \ln (a/a_i)$, with $a$ the scale factor of the Universe and $a_i$ its initial value, and the total equation of state $w_{tot} = p_{tot}/\rho_{tot}$. For $\lambda=0$ in Eq.~\eqref{eq:new4} the dynamical system for the free case is recovered; see Ref.~\cite{Urena-Lopez:2015gur}.

One critical step in the numerical solution of Eqs.~\eqref{eq:2} and~\eqref{eq:new4} is to find the correct initial conditions of the dynamical variables. For the axion case, it can be shown that we must satisfy the following constraints,
\begin{eqnarray}
\Omega_{\phi i} = a^{-3}_{\rm osc} a_i \frac{\Omega_{\phi 0}}{\Omega_{r 0}} \, , \quad y_{1i} = 5 \theta_i \, , \quad \frac{m_\phi^2}{H^2_i} = \frac{y^2_{1i}}{4} + \lambda \Omega_{\phi i} \, , \label{eq:initial}
\end{eqnarray}
where $a_{\rm osc}$ is the value of the scale factor at the onset of the oscillations of the field $\phi$ around the minimum of the potential~\eqref{eq:0}. The solution of Eqs.~\eqref{eq:initial} provides appropriate seed values that the CLASS code adjusts through a shooting procedure to obtain the correct value of the axion density parameter $\Omega_{\phi 0}$ at the present time.

\begin{figure}[tp!]
\includegraphics[width=0.48\textwidth]{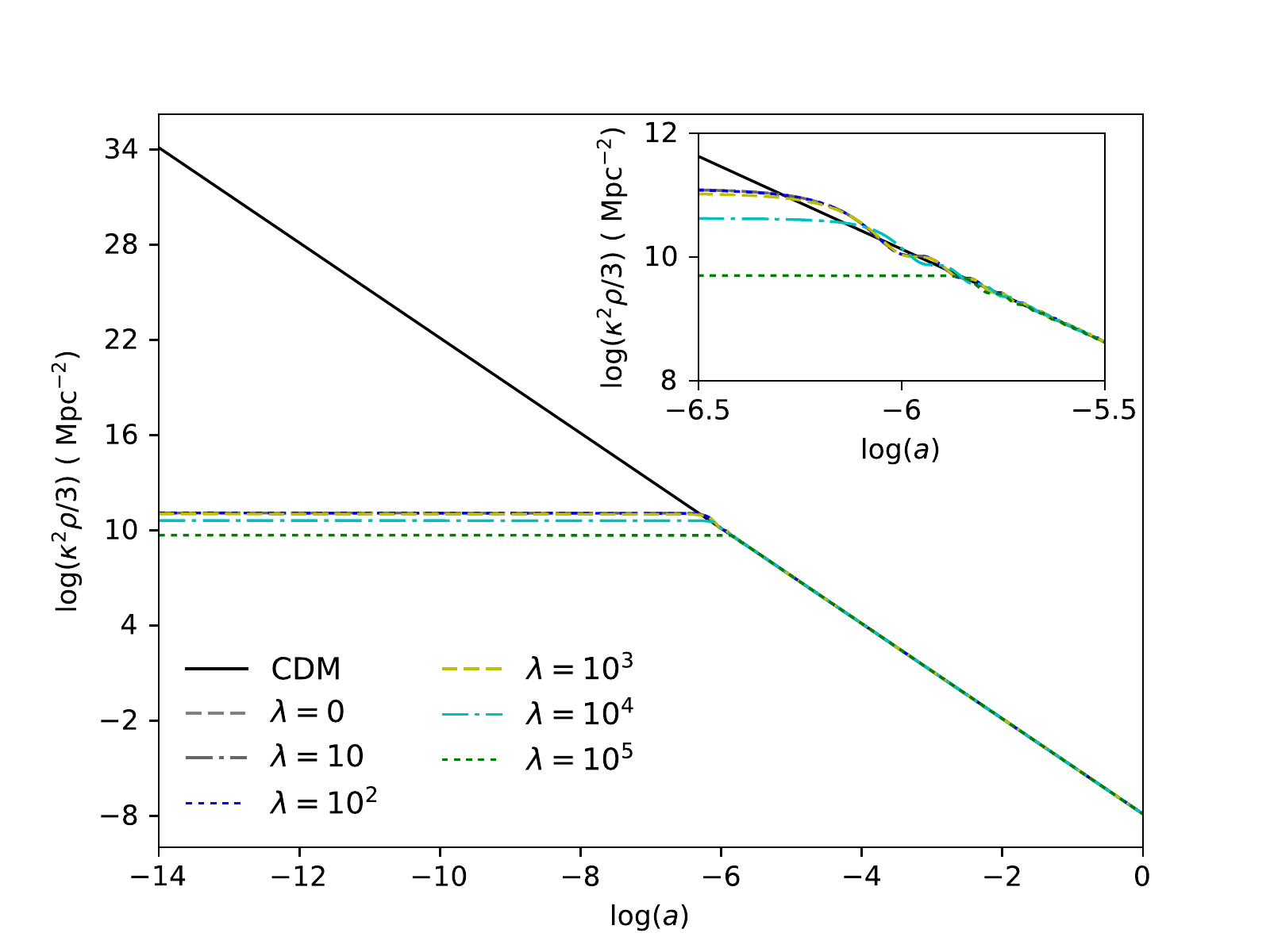}
\caption{The evolution of the axion and CDM energy densities up to the present time. Although the amplitude of the axion density is initially much smaller than that of CDM, it can be noted that from log$(a)\sim -6$ the axion density evolves exactly like CDM. The inset shows that for larger values of $\lambda$ the axion oscillations start later as compared to those of the free case, and that the transition to CDM happens more abruptly.}
\label{d}
\end{figure}

In Fig.~\ref{d} we show the evolution of the axion energy density $\rho_\phi$ in comparison with that of CDM (all other cosmological quantities are the same as in the fiducial $\Lambda$CDM model\cite{Ade:2015xua}). The numerical examples correspond to $\lambda=0, 10, 10^2, 10^3, 10^4, 10^5$. We can clearly see that $\rho_\phi$ evolves just like CDM after the onset of the field oscillations. The latter are delayed by the presence of the decay parameter $\lambda$, and also the transition to the CDM behavior occurs more abruptly  for larger values of $\lambda$. This is just a consequence of the increase in the steepness of the potential~\eqref{eq:0} for $\lambda \gg 1$, which in turn makes it more difficult to find a reliable numerical solution of Eqs.~\eqref{eq:new4}. The largest value considered for the decay parameter was $\lambda = 10^5$. Although larger values would be desirable, we are already close to the expected upper bound on $\lambda$. As estimated in Ref.~\cite{Diez-Tejedor:2017ivd}, the axion field can provide the whole of the DM budget as long as $m_\phi/\sqrt{\lambda} > 6 \times 10^{-27}$ eV. In particular, a conservative estimate is that $\lambda \lesssim 10^8$ if $m_\phi = 10^{-22}$ eV, although other considerations can provide stronger constraints~\cite{Diez-Tejedor:2017ivd,Visinelli:2017imh,Arias:2012az}.

\section{Linear Density Perturbations and Mass Power Spectrum \label{sec:linear-den}} 
Let us now consider the case of linear perturbations $\varphi$ of the axion field in the form $\phi(x,t) = \phi(t)+\varphi(x,t)$. As for the metric, we choose the synchronous gauge with the line element $ds^2 = -dt^2+a^2(t)(\delta_{ij}+h_{ij})dx^idx^j$, where $h_{ij}$ is the tensor of metric perturbations. The linearized Klein-Gordon equation for a given Fourier mode $\varphi(k,t)$ reads \cite{Ratra:1990me,Ferreira:1997au,Ferreira:1997hj,Perrotta:1998vf}:
\begin{equation}
  \ddot{\varphi} = - 3H \dot{\varphi} - \left[\frac{k^2}{a^2} -m_\phi^2\cos(\phi/f)\right] \varphi -
  \frac{1}{2} \dot{\phi} \dot{\bar{h}} \, , \label{eq:13}
\end{equation}
where a dot means derivative with respect the cosmic time, $\bar{h} = {h^j}_j$ and $k$ is a comoving wavenumber.

As shown in Ref.\cite{Urena-Lopez:2015gur}, we can transform Eq.~\eqref{eq:13} into a dynamical system by means of the following (generalized) change of variables,
\begin{equation}
\sqrt{\frac{2}{3}} \frac{\kappa \dot{\varphi}}{H} \equiv - \Omega^{1/2}_\phi e^{\alpha} \cos(\vartheta/2) \, , \; 
\frac{\kappa y_1 \varphi}{\sqrt{6}} \equiv - \Omega^{1/2}_{\phi} e^{\alpha} \sin(\vartheta/2) \, , \label{eq:linearvars}
\end{equation}
with $\alpha$ and $\vartheta$ the new variables needed for the evolution of the scalar field perturbations. But if we further define $\delta_0 = - e^{\alpha}\sin(\theta/2-\vartheta/2)$ and $\delta_1=-e^{\alpha}\cos(\theta/2-\vartheta/2)$, then Eq.~\eqref{eq:13} takes on a more manageable form,
\begin{subequations}
\label{eqdeltas}
\begin{eqnarray}
\delta^\prime_0 &=&  \left[-3\sin\theta-\frac{k^2}{k^2_J}(1 - \cos \theta) \right] \delta_1 + \frac{k^2}{k^2_J} \sin \theta \, \delta_0 \nonumber \\
&& - \frac{\bar{h}^\prime}{2}(1-\cos\theta) \, , \label{d0p} \\
\delta^\prime_1 &=& \left[-3\cos \theta - \left( \frac{k^2}{k^2_J} - \frac{\lambda \Omega_\phi}{2y_1} \right) \sin\theta \right] \delta_1 \nonumber \\
&+& \left(\frac{k^2}{k^2_J} - \frac{\lambda \Omega_\phi}{2y_1} \right) \left(1 + \cos \theta \right) \, \delta_0 - \frac{\bar{h}^\prime}{2} \sin \theta \, , \label{d1}
\end{eqnarray}
\end{subequations}
where $k_J^2 \equiv a^2 H^2 y_1$ is the (squared) Jeans wavenumber and a prime again denotes derivative with respect to the number of $e$-folds, $N$. Notice that the new dynamical variable
$\delta_0$ is the axion density contrast, as a straightforward calculation using Eqs.~\eqref{eq:backvars} and~\eqref{eq:linearvars} shows that $\delta \rho_\phi/\rho_\phi = (\dot{\phi}\dot{\varphi} + \partial_\phi V \varphi)/\rho_\phi =\delta_0$. This implies that Eq.~\eqref{d0p} is the closest expression one can find to a fluid equation for the evolution of the axion density contrast. The physical interpretation of $\delta_1$ is by no means as direct as that of $\delta_0$, and then Eq.~\eqref{d1} tells us of the difficulties to match the equations of motion of scalar field linear perturbations to those of a standard fluid\cite{Hu:1998kj}. For the initial conditions, we use the attractor solutions at early times\cite{Urena-Lopez:2015gur} given by $\delta_{0i} = - \bar{h}_i \theta^2_i/84$ and $\delta_{1i} = - \bar{h}_i \theta_i/7$, where $\bar{h}_i$ and $\theta_i$ are, respectively, the initial values of the trace of metric perturbations $\bar{h}$ and the background angular variable $\theta$.

\begin{figure}[tp!]
\includegraphics[width=0.48\textwidth]{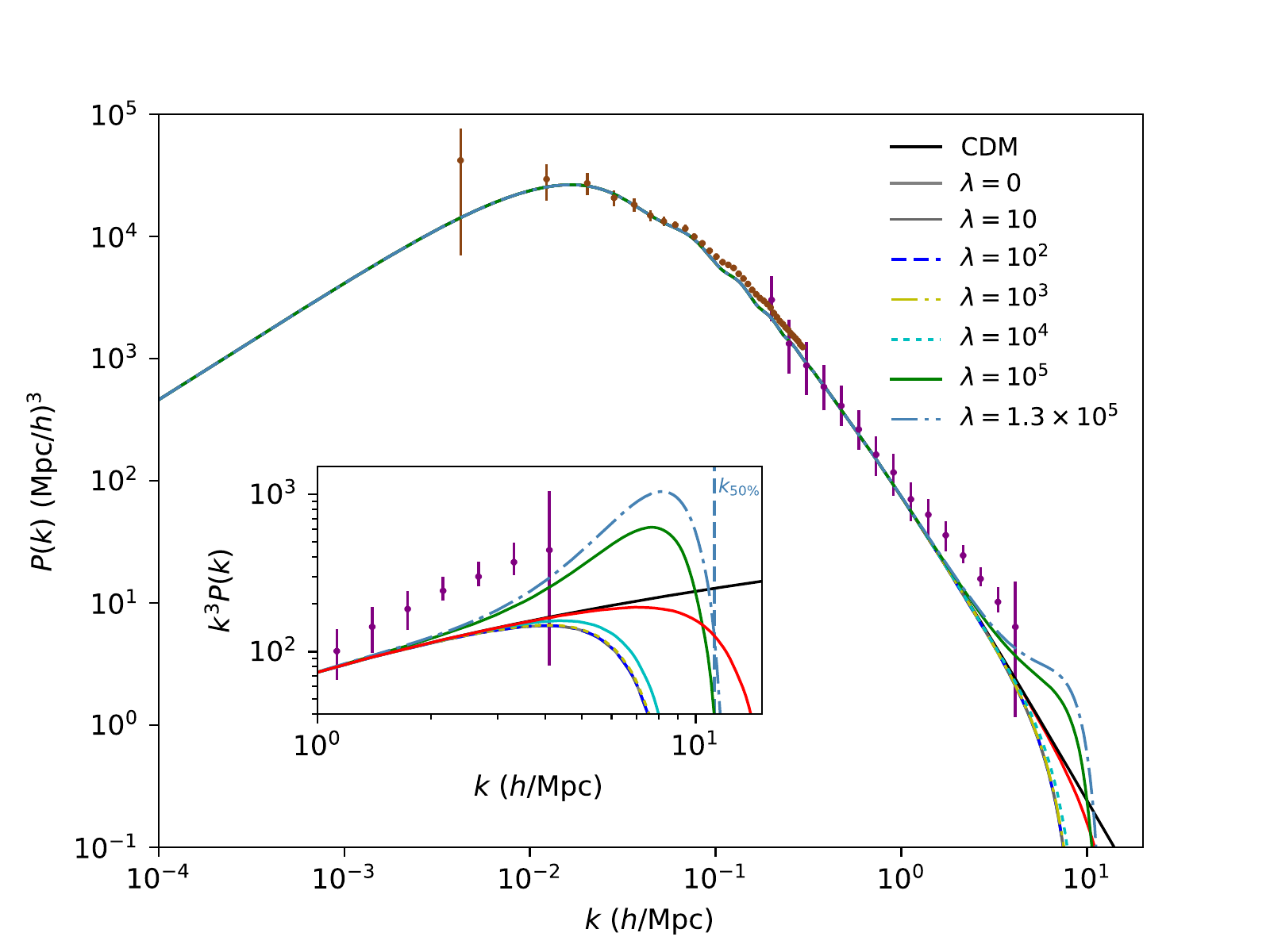}
\caption{MPS at the present time for an axion field with the same values of $\lambda$ as in Fig.\ref{d}. The characteristic cut-off of the axion MPS is clearly seen, together with some differences at small scales induced by the physical parameter, $\lambda$. MPS data from BOSS DR11 (brown dots) \cite{Anderson:2013zyy} and from Ly$\alpha$ forest (purple dots) \cite{Zaroubi:2005xx} are shown for reference. (Inset) Zoom in of the MPS for large values of $k$. It is concluded that the presence of the parameter $\lambda$ produces a bump when $\lambda\gg 1$. The blue dashed vertical line shows the difference between the extreme case $\lambda=1.3\times 10^5$ with respect to CDM at a $50\%$. The red line represents the free case with a mass $m_\phi^*=3.635\times 10^{-22}$eV, whose MPS differs at 50\% from CDM just as the extreme axion case, at a wavenumber, $k_{50\%}=11.218h/$Mpc. See the text for more details.}
\label{ps22}
\end{figure}

The solution of Eqs.~\eqref{eqdeltas} are useful to build up cosmological observables such as the mass power spectrum (MPS), which we show for the axion field and CDM in Fig.~\ref{ps22}. It is well known that there is a characteristic cut-off in the MPS of a free field, and this feature is also present for the axion case, although the cut-off is shifted towards smaller scales (larger wavenumbers). But more prominently, the axion MPS presents an excess of power, even compared to CDM, at scales close to the cut-off if $\lambda \gg 1$. Such excess was reported before in Refs.\cite{Zhang:2017flu,Zhang:2017dpp} (see also Ref.~\cite{Suarez:2011yf} for an early indication of such power excess in scalar field models) and attributed to the so-called extreme initial conditions of the background field (under our approach, this means $\phi/f \to 0$). As we shall explain below, the excess should be rightfully attributed to the extreme value of $\lambda \gg 1$ [which in turn has an effect on the initial conditions via Eqs.~\eqref{eq:initial}], and then ultimately to the decay constant $f$.

Also shown in Fig.~\ref{ps22} are the free (with mass $m_\phi^*=3.635\times 10^{-22}$eV and $\lambda=0$) and extreme cases (with $\lambda=1.3\times 10^5$) whose MPS differs by $50\%$ from that of CDM at the same wavenumber, namely $k_{50\%}=11.218h/$Mpc. However, it is important to highlight that both cases have a very different behavior at smaller and larger values of $k$, which means that the axion MPS is non-degenerate with respect to that of the free case. Moreover, this also shows that the axion case $(m_\phi,\lambda\neq 0)$ cannot be exactly matched to a free case $(m_\phi^*,\lambda= 0)$. 

As for the excess of power at some scales in the MPS, we first note that the presence of $\lambda$ in Eq.~\eqref{d1} defines an effective wavenumber in the form $k^2_{eff} = k^2 - \lambda a^2 H^2 \Omega_\phi/2$, which, in contrast to the ratio $k^2/k^2_J$ that appears in the free case, it could be positive as well as negative. Taking advantage of the similarities with the free case\cite{Urena-Lopez:2015gur}, we will study the homogeneous solutions of Eqs.~\eqref{eqdeltas} (without the driving terms) after the onset of the rapid oscillations of the axion field. We first discard all the trigonometric terms, and then Eq.~\eqref{eqdeltas} can be written as: $\delta^\prime_0 \simeq - (k^2/k^2_J) \delta_1$ and $\delta^\prime_1 \simeq (k^2_{eff}/k^2_J) \delta_0$. Under the assumption that both functions $k^2_{eff}$ and $k^2_J$ are approximately constant, we obtain that the density contrast has a harmonic solution of the form $\delta_0 \sim C_0 \cos(\omega N)$, where the (squared) fundamental frequency is $\omega^2 = k^2 k^2_{eff}/k^2_J$ and $C_0$ is an integration constant. Just like in the free case, it can be seen that if $0 < |\omega^2| \ll 1$ the homogeneous solution of Eqs.~\eqref{eqdeltas} becomes irrelevant and then the axion density contrast can grow similarly to that of CDM. Similarly, if $1 \ll \omega^2$ the growth of the density contrast is strongly suppressed and there appears a sharp cut-off in the MPS at large $k$ (small scales). But now we must also consider the possibility that $\omega^2 \ll -1$, for which the homogeneous solution changes to $\delta_0 \sim C_0 \cosh(|\omega|N)$, and then the growth of the given mode $k$ is even enhanced beyond the CDM case. We dub the latter effect as the \textit{tachyonic instability} of linear perturbations.

\begin{figure}[tp!]
\includegraphics[width=0.48\textwidth]{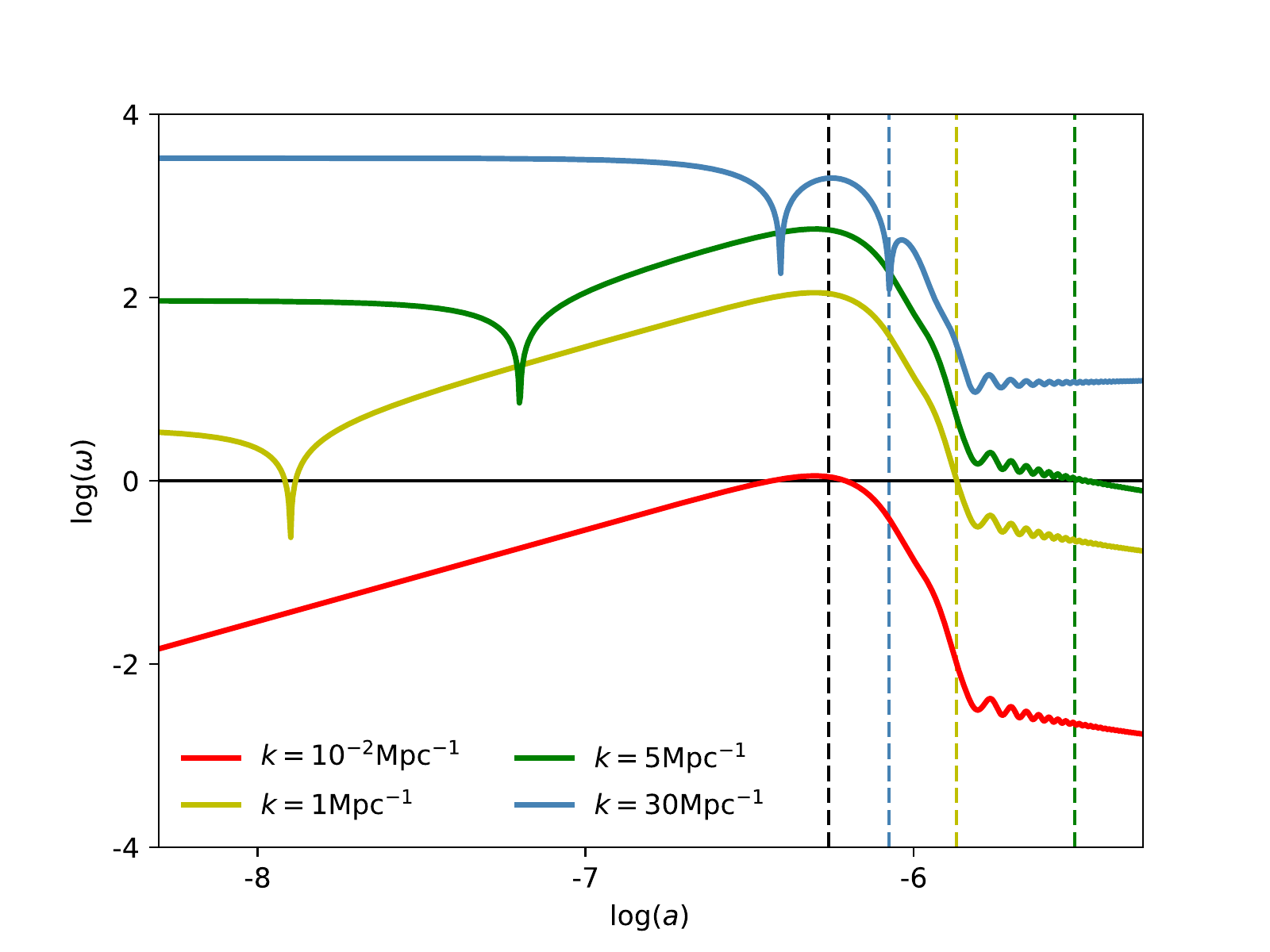}
\caption{Evolution of the amplitude of the (complex) frequency $\omega = k |k_{eff}|/k_J^2$ that is intrinsic to the system of linear perturbations~\eqref{eqdeltas}, for different wavenumbers $k$ but the same $\lambda =10^5$. We can see that the frequency becomes purely imaginary, $\omega^2 < 0$, for all the modes before the onset of the axion oscillations (marked by the black dashed vertical line), but it is just for a handful of them that a tachyonic instability ($\omega^2 \ll -1$) lasts for long enough: $10^{-2} < k/ \textrm{Mpc}^{-1} < 30$. For reference, the end of the instability for each mode in the plot is indicated by the vertical dashed lines of the same color. As a result, large scales must evolve like CDM, and in the MPS of Fig.~\ref{ps22} there must be a bump for the aforementioned range of wavenumbers together with a displacement of the cut-off towards smaller scales when compared to the free case.}
\label{kj}
\end{figure}

To determine the linear modes that suffer a tachyonic instability we proceed as follows. We note that both the Jeans wavenumber $k_J$ and the effective wavenumber $k_{eff}$ are functions of background quantities only, and then their evolution can be easily calculated for different values of $\lambda$ and $k$. This is shown in Fig.~\ref{kj}, where we see that only a limited range of $k$, and for a finite lapse of time after the onset of the rapid oscillations of the axion field, will be affected by the tachyonic instability $\omega^2 \ll -1$. The wavenumbers shown in Fig.~\ref{kj} constitute a representative set of modes that allow us to get a better comprehension of the bump in the MPS for the extreme case $\lambda = 10^5$. Large scales with $k < 10^{-2}$Mpc$^{-1}$ are not affected by the tachyonic instability because for them the condition $\omega^2 \ll -1$ is never satisfied. Modes with $k > 10^{-2}$Mpc$^{-1}$ start to be affected as the condition $\omega^2 \ll -1$ is satisfied at the onset of the oscillations of the axion field, but the time lapse of the tachyonic effect is reduced as the wavenumber increases (in other words, it takes less and less time for $\omega^2$ to change from negative to positive again), so that for small scales with $k > 30$Mpc$^{-1}$ the tachyonic instability never happens. In fact, $\omega^2 \gg 1$ at all times for those latter modes and the result is that the amplitude of their perturbations must be highly suppressed. Thus, we infer from Fig.~\ref{kj} that the tachyonic instability only affects the modes with $10^{-2} < k/ \textrm{Mpc}^{-1} < 30$, which includes the range of wavenumbers responsible for the bump in the MPS in Fig.~\ref{ps22}, i.e., approximately $1 < k/ \textrm{Mpc}^{-1} < 10$.

\section{Discussion and conclusions \label{sec:discussion-and}}
We fully computed the MPS for the axion potential and, within the range of physical parameters that we were able to explore, showed that its features do change significantly in the case $f/m_{Pl} \ll 1$ ($\lambda \gg 1$), for which linear perturbations in a certain range of wavenumbers suffer a tachyonic instability and are able to grow more than their CDM counterparts. This causes the appearance of a bump in the MPS which is close to the cut-off scale, which in turn is also displaced towards larger wavenumbers in comparison to the free case. Our results are in agreement with the semi-analytical studies of the axion field in Refs.~\cite{Zhang:2017flu,Zhang:2017dpp,Suarez:2011yf} (see also~\cite{Diez-Tejedor:2017ivd}), which were the first to suggest the existence of a bump in the MPS. However, we were able to show that such effect results from the condition $\lambda \gg 1$\cite{Suarez:2011yf}, rather than from extreme initial condition $\phi_i/f \ll 1$ as suggested in Refs.~\cite{Zhang:2017flu,Zhang:2017dpp}. 

\begin{figure}[tp!]
\includegraphics[width=0.48\textwidth]{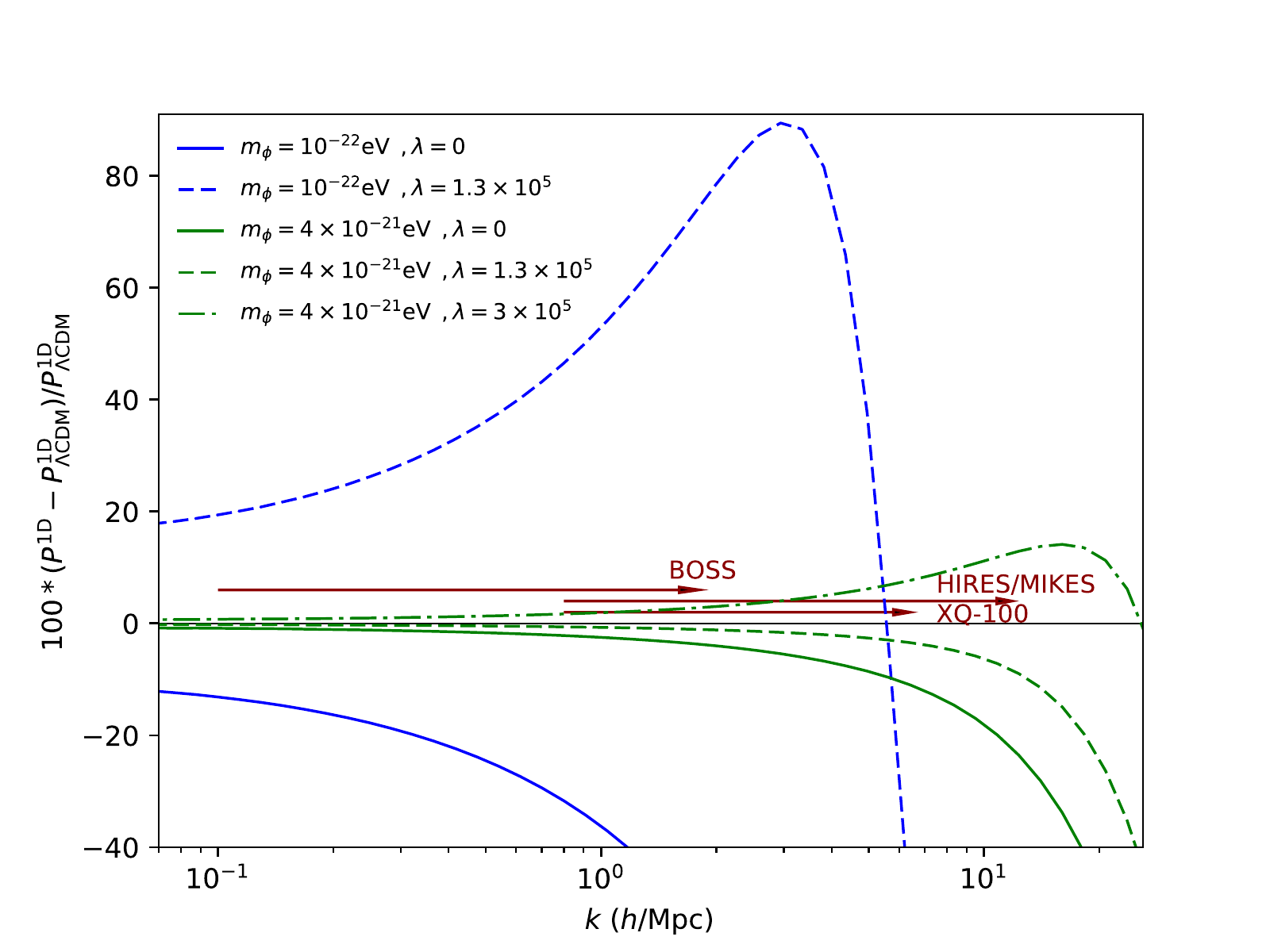}
\caption{1D MPS for the Axion field compared to the $\Lambda$CDM one. We show the cases $m_{\phi}=10^{-22} {\rm eV}$ for $\lambda=0,1.3 \times 10^5$ (blue lines), and $m_{\phi}=4\times 10^{-21} {\rm eV}$ for $\lambda=0,1.3 \times 10^5, 3 \times 10^5$ (green lines). For reference we have included  horizontal lines indicating the rough precision of current data from BOSS \cite{Palanque-Delabrouille:2013gaa}, HIRES/MIKES\cite{Viel:2013apy} and XQ100\cite{2016A&A...594A..91L} to show that this can be used to constraint both parameters $m_\phi$ and $\lambda$. }
\label{fig:ps1d}
\end{figure}

Just recently a set of new constraints on the axion mass based on the analysis of Lyman-$\alpha$ forest had been presented. The strongest constraint comes from high resolution spectra, implying $m_\phi > 37.5\times 10^{-22}$eV ~\cite{Irsic:2017yje} and $m_\phi > 29 \times 10^{-22}$eV \cite{Armengaud:2017nkf}, at the 2-$\sigma$ confidence level. To extrapolate such constraints to the full axion potential is not straightforward. In Fig.~\ref{fig:ps1d} we show the 1-dimensional MPS ($P^{1{\rm D}}$) for the full axion potential, relative to that of  the $\Lambda$CDM. The $P^{1{\rm D}}$ is closely related to the flux power spectrum, $P_{\rm F}$, that is the actual observable in surveys like BOSS \cite{Palanque-Delabrouille:2013gaa}, HIRES/MIKES\cite{Viel:2013apy} and XQ100\cite{2016A&A...594A..91L}. For reference we have included horizontal lines at the approximate precision at which such experiments can prove the $P^{1{\rm D}}$, and the $k$-range they cover. This can be read as indication that such observations would be able to set constraints in the axion mass vs decay parameter ($m_\phi$ vs $\lambda$) plane, provided that the actual quantity one have to predict is  $P_{\rm F}$ following an analysis similar to that in~\cite{Irsic:2017yje,Armengaud:2017nkf}. It will also be interesting to know whether future surveys as the Dark Energy Spectroscopic Instrument (DESI)\cite{Levi:2013gra} and the Large Synoptic Survey Telescope (LSST)\cite{Ivezic:2008fe}, in case a bump and a cut-off in the MPS, or the $P^{1{\rm D}}$ are detected, will also be able to spot the differences between the free case and the full axion one and in turn put constraints on the decay parameter $\lambda$.

Some discussion about the consequences of a so-called extreme axion case in the formation of cosmological structure has been put forward in Ref.~\cite{Schive:2017biq}, by considering the equivalence between N-body simulations and the Schrodinger-Poisson system first hinted at in Ref.~\cite{Schive:2014dra}. From our perspective, the formation of structure under the extreme axion case is better captured by a Gross-Pitaevskii-type of equation with a negative-definite quartic self-interaction given by $g_4 = - m^2_\phi f^{-2}/6$ (the coefficient $g_4$ can be read off from the series expansion of potential~\eqref{eq:0} up to the fourth order: $V \simeq (m^2_\phi/2) \phi^2 - (m^2_\phi f^{-2}/4!) \phi^4$. According to diverse studies in Refs.~\cite{Guzman:2006yc,Chavanis:2011zi,Levkov:2016rkk}, the parameter that quantifies the strength of the self-interaction is the combination $g_4/(4 \pi G m^2_\phi) = - \lambda/3$. In the extreme case $\lambda \gg 1$, equilibrium gravitational configurations of the coupled Gross-Pitaevskii-Poisson system present stable and unstable branches (see also Refs.~\cite{Barranco:2013wy,Helfer:2016ljl} for the relativistic axion case), and the critical quantities at the transition point between the two branches have been found to be $\phi_c/f \simeq \lambda^{-1/2}$ (for the central field value) and $M_c \simeq \lambda^{-1/2} \, m^2_{\rm Pl}/m_\phi$ (for the total mass), where $m_{\rm Pl}$ is the Planck mass~\cite{Guzman:2006yc,Chavanis:2011zi,Levkov:2016rkk}. On one hand, stable configurations then correspond to field values $\phi/f \leq \lambda^{-1/2}$, and then the gravitational stability of an axion configuration requires a more diluted field for larger $\lambda$. At the same time, the critical total mass $M_c$ also decreases for larger $\lambda$, and for the fiducial model considered throughout we find $M_c \sim 10^9 \, M_\odot$ if $\lambda = 10^5$. As already noted in Refs.~\cite{Levkov:2016rkk}, this means that even the less massive halo objects in a typical simulation (see for instance~\cite{Schive:2014dra}) would be in risk to collapse into black holes.

All of the above lead us to wonder about the possibility of having $\lambda < 0$, so that the quartic self-interaction $g_4$ is strictly positive definite. In such a case, the gravitational stability of bounded objects is instead enhanced by the presence of $\lambda$ and then the difficulties of the extreme axion case are easily avoided~\cite{Colpi:1986ye,UrenaLopez:2001tw,Liebling:2012fv,Herdeiro:2016gxs,UrenaLopez:2012zz,Li:2013nal,RindlerDaller:2011kx,Rindler-Daller:2013zxa,Diez-Tejedor:2014naa}. This requires, at least formally, that $(f/m_{\rm Pl})^2 <0$ and then the trigonometric potential~\ref{eq:0} is replaced by the hyperbolic one studied in Refs.~\cite{Matos:2000ng,Matos:2000ss,Sahni:1999qe}. The study of such case is part of ongoing work that will be presented elsewhere.

\section*{Acknowledgements} We would like to thank Alberto Diez-Tejedor for useful discussions and comments. FXLC acknowledges CONACYT for financial support. AXG-M acknowledges support from C\'atedras CONACYT and UCMEXUS-CONACYT collaborative project funding. This work was partially supported by Programa para el Desarrollo Profesional Docente; Direcci\'on de Apoyo a la Investigaci\'on y al Posgrado, Universidad de Guanajuato, research Grants No. 732/2017 y 878/2017; CONACyT M\'exico under Grants No. 167335, No. 179881, No. 269652 and Fronteras 281; and the Fundaci\'on Marcos Moshinsky.

\section*{Bibliography}
\bibliography{bib}

\end{document}